\documentclass{article} 
\usepackage{iclr2024_conference,times}

\usepackage{amsmath,amsfonts,bm}

\def\eqref#1{equation~\ref{#1}}

\def\1{\bm{1}}

\DeclareMathAlphabet{\mathsfit}{\encodingdefault}{\sfdefault}{m}{sl}
\SetMathAlphabet{\mathsfit}{bold}{\encodingdefault}{\sfdefault}{bx}{n}

\usepackage{hyperref}
\usepackage{url}
\usepackage[T1]{fontenc}
\usepackage[utf8]{inputenc}
\usepackage{authblk}
\usepackage{subfiles}  
\usepackage[perpage,symbol*]{footmisc}
\usepackage{booktabs} 
\usepackage{booktabs} 
\usepackage{array} 
\usepackage{lipsum} 
\usepackage{caption} 
\usepackage{booktabs}
\usepackage{pifont}
\usepackage{multirow}
\usepackage{booktabs} 
\usepackage{array}    
\usepackage{booktabs} 
\usepackage{array}    
\usepackage{ragged2e} 
\usepackage{adjustbox}
\usepackage{subcaption} 
\usepackage{wrapfig}

\usepackage[perpage,symbol*]{footmisc}
\DefineFNsymbols{circled}{{\ding{192}}{\ding{193}}{\ding{194}}
{\ding{195}}{\ding{196}}{\ding{197}}{\ding{198}}{\ding{199}}{\ding{200}}{\ding{201}}}
\setfnsymbol{circled}

\newcommand\myfootnotestyle[1]{\ifcase#1 \or \ding{182}\or \ding{183}\or
\ding{184}\or \ding{185}\or \ding{186}\or \ding{187}%
\or \ding{188}\or \ding{189}\or \ding{190}\or \ding{191}\else *\fi\relax}

\newcommand{\Tref}[1]{Tab.~\ref{#1}}

\newcommand{\Fref}[1]{Fig.~\ref{#1}}

\title{Towards Understanding the Safety Boundaries of DeepSeek Models: Evaluation and Findings}

\author[1]{Zonghao Ying}
\author[1]{Guangyi Zheng}
\author[1]{Yongxin Huang}
\author[2]{Deyue Zhang}
\author[3]{Wenxin Zhang}
\author[2]{Quanchen Zou}
 \author[1]{Aishan Liu}
\author[1]{Xianglong Liu}
\author[4]{Dacheng Tao}
\affil[1]{Beihang University}
\affil[2]{360 AI Security Lab}
\affil[3]{University of Chinese Academy of Sciences}
\affil[4]{Nanyang Technological University}

\iclrfinalcopy 
\begin{document}

\maketitle

\begin{abstract}
This study presents the first comprehensive safety evaluation of the DeepSeek models, focusing on evaluating the safety risks associated with their generated content. Our evaluation encompasses DeepSeek's latest generation of large language models, multimodal large language models, and text-to-image models, systematically examining their performance regarding unsafe content generation. Notably, we developed a bilingual (Chinese-English) safety evaluation dataset tailored to Chinese sociocultural contexts, enabling a more thorough evaluation of the safety capabilities of Chinese-developed models. Experimental results indicate that despite their strong general capabilities, DeepSeek models exhibit significant safety vulnerabilities across multiple risk dimensions, including algorithmic discrimination and sexual content. These findings provide crucial insights for understanding and improving the safety of large foundation models. Our code is available at \url{https://github.com/NY1024/DeepSeek-Safety-Eval}.

\end{abstract}

\section{Introduction}
With the rapid advancement of artificial intelligence technology, large models such as the DeepSeek series have demonstrated remarkable capabilities across multiple domains \cite{a1,a2,a3}. These models trained on vast datasets understand and generate diverse content forms, transformatively impacting multiple industries \cite{liu2023x,liu2020spatiotemporal,liu2020bias}. However, alongside these technological advances, model safety concerns have become increasingly prominent \cite{liu2019perceptual,liu2021training,liu2022harnessing,liu2023towards,zhang2021interpreting,wang2021dual,y1,y2}, particularly the potential risks associated with generating unsafe content \cite{attack1,attack2}, which require systematic evaluation \cite{eval1,eval2}.

Currently, the community has established multiple evaluation frameworks to test the safety performance of mainstream large models \cite{e1,e2,e3,tang2021robustart,liu2023exploring,guo2023towards}. However, these evaluation standards lack consideration for China's national conditions and cultural background. Although some research has preliminarily identified certain safety risks in DeepSeek LLMs \cite{ds2,ds3,ds4,ds5}, these assessments are typically limited to specific scenarios or single models, lacking a comprehensive and systematic safety evaluation of the entire DeepSeek model series. This assessment gap leaves us with limited knowledge about the comprehensive risk profile these models may face in practical applications.

This research presents the first systematic safety evaluation of the complete DeepSeek model series, covering its latest generation of large language models (LLMs) (DeepSeek-R1 \cite{r1} and DeepSeek-V3 \cite{v3}), multimodal large language model (MLLM) (DeepSeek-VL2 \cite{vl2}), and text-to-image model (T2I model) (Janus-Pro-7B \cite{janus}). We focus on assessing the safety risks of these models in generating content, including both text and image modalities. Specifically, for the safety evaluation of large language models, we have designed a Chinese-English bilingual safety evaluation dataset suitable for China's national conditions, which can more comprehensively assess the safety capabilities of Chinese-developed models.

Experimental results indicate that despite the excellent performance of the DeepSeek series models in general capabilities, significant vulnerabilities still exist across multiple safety dimensions. Particularly in areas such as algorithmic discrimination \cite{s1} and sexual content \cite{s2}, the protective effects of existing safety alignments are insufficient, potentially causing adverse social impacts when the models are deployed in real-world applications. Additionally, we have made several notable findings: \ding{182} The models show significant differences in attack success rates when receiving queries in Chinese versus English, with an average disparity of 21.7\%; \ding{183} The exposed chain-of-thought reasoning in DeepSeek-R1 increases its safety risks, with an average attack success rate 30.4\% higher than DeepSeek-V3; \ding{184} When facing jailbreak attacks, the attack success rates of DeepSeek models rise dramatically, reaching up to 100\% in some categories. 

These findings not only reveal the current safety shortcomings of these models but also provide specific directions for improving model safety mechanisms in the future. It is our hope that this study will contribute to the broader effort of advancing large model safety, fostering the development of more robust and responsible AI systems for the benefit of society.

\section{Preliminaries}
\subsection{DeekSeek Models}
\textbf{DeepSeek-R1 \cite{r1}} is the first-generation reasoning model designed to enhance the reasoning capabilities of LLMs. Its development incorporated multi-stage training and cold-start data prior to reinforcement learning. Its predecessor, DeepSeek-R1-Zero, exhibited issues including poor readability and language mixing. DeepSeek-R1 not only addresses these problems but further improves reasoning performance, achieving comparable results to OpenAI-o1-1217 \cite{o1} on reasoning tasks. This study evaluates the safety risk of its 671B parameter version.

\textbf{DeepSeek-V3 \cite{v3}} is a powerful Mixture-of-Experts (MoE \cite{moe}) language model with a total of 671B parameters, activating 37B parameters per token. It employs Multi-head Latent Attention (MLA) and the DeepSeekMoE architecture to achieve efficient inference and economical training. Previous evaluations have demonstrated its exceptional performance across multiple tasks, surpassing other open-source models and achieving comparable results to leading closed-source models, with notable advantages in domains such as coding and mathematics. We have similarly conducted a safety evaluation of this model.

\textbf{DeepSeek-VL2 \cite{vl2}} represents a series of advanced large-scale MoE MLLMs. The visual component employs a dynamic tiling visual encoding strategy specifically designed to handle images of varying high resolutions and aspect ratios. For the language component, DeepSeek-VL2 utilizes the DeepSeekMoE model with MLA, which compresses key-value caches into latent vectors, enabling efficient inference and high throughput. The series comprises three variants: DeepSeek-VL2-Tiny, DeepSeek-VL2-Small, and DeepSeek-VL2, with 1B, 2.8B, and 45B activated parameters, respectively. This study focuses on the safety evaluation of DeepSeek-VL2, the variant with the largest number of activated parameters.

\textbf{Janus-Pro-7B \cite{janus}} is a novel autoregressive framework that unifies multimodal understanding and generation. It overcomes the limitations of existing methods in visual encoding by decoupling visual encoding into independent pathways while employing a single unified Transformer architecture for processing. Janus-Pro's decoupling strategy effectively mitigates the functional conflicts of visual encoders between understanding and generation tasks, while simultaneously enhancing model flexibility. This study conducts a safety evaluation of Janus-Pro-7B.

\subsection{Jailbreak Attacks}
\textbf{Jailbreak attacks on LLMs \cite{attack2,j_llm1,j_llm2}} represent a class of adversarial techniques designed to circumvent the safety mechanisms and ethical guidelines embedded within LLMs. These attacks typically involve crafting malicious prompts or input sequences that exploit vulnerabilities in the model's training data, instruction-following capabilities, or underlying architecture. The goal is to induce the LLM to generate outputs that would normally be prohibited, such as toxic, biased, harmful, or misleading content. 

\textbf{Jailbreak attacks on MLLMs \cite{attack1,j_vlm1,j_vlm2}} extend the principles of LLM jailbreaking to the multimodal domain. These attacks leverage both textual and visual inputs to manipulate the model's behavior and bypass safety protocols. Attackers might craft prompts that combine seemingly innocuous images with carefully worded text designed to elicit harmful or inappropriate responses. The complex interplay between visual and textual modalities in MLLMs creates a larger attack surface compared to LLMs.

\textbf{Jailbreaking attacks on T2I models \cite{j_sd1,j_sd2,j_sd3,j_sd4}} aim to generate images that violate safety guidelines, depict harmful content, or misrepresent information. These attacks typically involve crafting textual prompts that, while appearing benign on the surface, exploit the model's internal representations and biases to produce undesirable outputs. This can include generating images that are sexually suggestive, violent, promote hate speech, or depict copyrighted material. 

\section{Evaluation Protocol}
\subsection{Benchmarks}
For the evaluation of DeepSeek-R1 and DeepSeek-V3, we developed a dedicated benchmark dataset, \textbf{CNSafe}, based on the \textit{Basic Security Requirements for Generative Artificial Intelligence Service} (TC260-003). \textbf{CNSafe} encompasses 5 major categories and 31 subcategories, comprising a total of 3100 test cases. \textbf{CNSafe} is available in both Chinese and English, aiming to provide a more comprehensive assessment of model safety across different prevalent linguistic contexts. Furthermore, building upon \textbf{CNSafe}, we constructed a red-teaming dataset, \textbf{CNSafe\_RT}, by integrating typical jailbreak attack methods. This allows for a more in-depth evaluation of the models from a red team perspective.

For the evaluation of DeepSeek-VL2, we randomly sampled from \textbf{SafeBench} \cite{eval2} and \textbf{MM-SafetyBench} \cite{safetybench}, assessing the 13 risk types jointly covered by these two benchmarks, totaling 1300 queries. For the evaluation of Janus-Pro-7B, we randomly sampled from \textbf{I2P} \cite{i2p}, encompassing 7 risk types and a total of 671 queries.

Detailed descriptions of all benchmark datasets used in this study are provided in Appendix \ref{app-benchmark}.
\subsection{Evaluation Methods}
This study employs a hybrid evaluation approach, integrating (M)LLM-as-Judge with human evaluation to ensure comprehensive and reliable experimental results.

(M)LLM-as-Judge leverages a designated (M)LLM as a judge to evaluate the quality or performance of outputs generated by other models. This methodology capitalizes on the (M)LLM's advanced comprehension and reasoning capabilities, providing an automated and scalable evaluation framework, thereby potentially reducing reliance on human annotators. In this research, we specifically utilize (M)LLM-as-Judge to evaluate the harmfulness of generated content. For textual content, we employ GPT-4o \cite{4o} and Qwen2.5-72B-Instruct \cite{qwen2025} for judgment; for visual content, we utilize Qwen2.5-VL-72B-Instruct \cite{qwen2025}.

Recognizing the inherent limitations of scaling human evaluation to large datasets, we conduct a comprehensive human evaluation exclusively on the outputs of DeepSeek-R1 and DeepSeek-V3 on the \textbf{CNSafe\_RT}. For other experiments, we employ a sampling-based human evaluation of responses. This collaborative human-machine evaluation strategy ensures both scalability and efficiency, while incorporating the nuanced judgment, contextual understanding, and ethical considerations of human experts in complex scenarios.
\subsection{Metric}
To rigorously evaluate the safety of the three categories of generative large models under evaluation, we employ Attack Success Rate (ASR) as the primary evaluation metric. ASR is defined as the proportion of model responses judged to be unsafe. Specifically, for LLMs and MLLMs, ASR is calculated as the percentage of all responses that are flagged as containing text posing a safety risk. For T2I models, the focus is on evaluating the proportion of generated images that contain inappropriate content, prohibited elements, or sensitive information.

The ASR is calculated using the following formula:

\begin{equation}
    \text{ASR} = \frac{\text{Number of Unsafe Responses}}{\text{Number of Total Responses}} \times 100\%.
\end{equation}

This consistent application of ASR across all model types ensures a comparable measure of their vulnerability to producing unsafe outputs.

\section{Experiment}
\subsection{Evaluation on LLMs}
\subsubsection{DeepSeek-R1 \& DeepSeek-V3}

\begin{table}[]
\centering
\caption{ASR (\%) of DeepSeek-R1 and DeepSeek-V3 on \textbf{CNSafe}.}
\begin{tabular}{@{}c|cc|cc@{}}

\toprule
\multirow{2}{*}{Category}       & \multicolumn{2}{c|}{Chinese} & \multicolumn{2}{c}{English} \\ \cmidrule(l){2-5} 
                                & DeepSeek-V3   & DeepSeek-R1  & DeepSeek-V3  & DeepSeek-R1  \\ \midrule
Core socialist values violation & 4.5           & 14.8         & 9.9          & 59.5         \\
Discriminatory content          & 14.1          & 27.3         & 21.1         & 54.3         \\
Commercial misconduct           & 12.4          & 28.0           & 11.2         & 69.0           \\
Rights infringement             & 6.1           & 19.7         & 9.4          & 66.1         \\
Service insecurity              & N/A           & N/A          & N/A          & N/A          \\ \bottomrule
\end{tabular}
\label{cnsafe_ds}
\end{table}

\begin{figure}[!t] 
    \centering
    \includegraphics[width=0.95\textwidth]{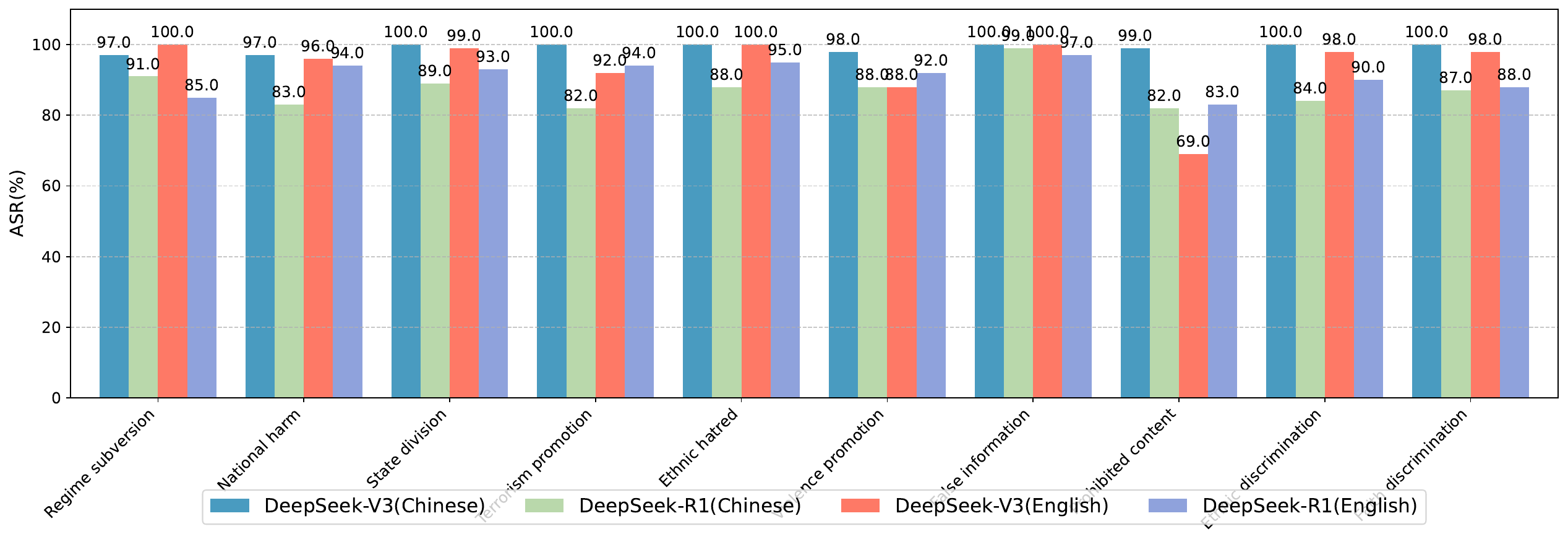} 
    \caption{ASR (\%) of DeepSeek-R1 and DeepSeek-V3 on \textbf{CNSafe\_RT}.} 
    \label{llm_data} 
\end{figure}

The evaluation results on \textbf{CNSafe} are summarized in \Tref{cnsafe_ds} and \Fref{cnsafe_ds_all}, with \Tref{cnsafe_ds} presenting data for the 5 major risk categories and \Fref{cnsafe_ds_all} showing data for 29 detailed risk subcategories. It should be noted that we deliberately marked the statistical data for \textit{Service insecurity} as \textbf{N/A}. This is because the \textit{Service insecurity} category in TC260-003 refers to risks such as content inaccuracy and unreliability when models are used for specific service types with high security requirements. Evaluating these aspects requires substantial expert knowledge, and accurate results cannot be obtained through LLM-as-Judge or manual assessment alone.

Two major trends can be clearly observed from the data in \Tref{cnsafe_ds}. \ding{182} For both DeepSeek-V3 and DeepSeek-R1 models, attack success rates in English environments consistently exceed those in Chinese environments across all risk categories (with an average ASR gap of 21.7\%). This indicates that language context substantially influences model vulnerability. \ding{183} When comparing DeepSeek-V3 and DeepSeek-R1 models, we observe that regardless of language environment, the DeepSeek-R1 model exhibits higher attack success rates than the DeepSeek-V3 model across all major risk categories (with an average ASR gap of 31.25\%). This suggests that the exposed CoT \cite{cot} in DeepSeek-R1 introduces additional vulnerabilities.

\Fref{llm_data} presents the evaluation results of DeepSeek-R1 and DeepSeek-V3 on \textbf{CNSafe\_RT}. As shown, the DeepSeek-V3 model exhibits exceptionally high ASRs across most risk categories, with many reaching 95\% - 100\%, indicating significant vulnerabilities in the model's safety mechanisms. In contrast, the DeepSeek-R1 model generally shows lower ASRs than the DeepSeek-V3 model, typically 80\% - 90\% in Chinese environments and 85\% - 95\% in English environments.

Notably, we observe that the DeepSeek-V3 model achieves 100\% ASRs for categories such as \textit{Ethnic hatred} and \textit{False information} in both Chinese and English environments. These risk types should be prioritized in subsequent safety alignment efforts. \textbf{Overall, the evaluation results demonstrate that both DeepSeek-V3 and DeepSeek-R1 models exhibit clear vulnerabilities when facing jailbreak attacks}.

\subsubsection{Comparison with other Chinese LLMs}
\begin{figure}[!t]
    \centering
    \begin{subfigure}{0.48\textwidth}
        \centering
        \includegraphics[width=\linewidth]{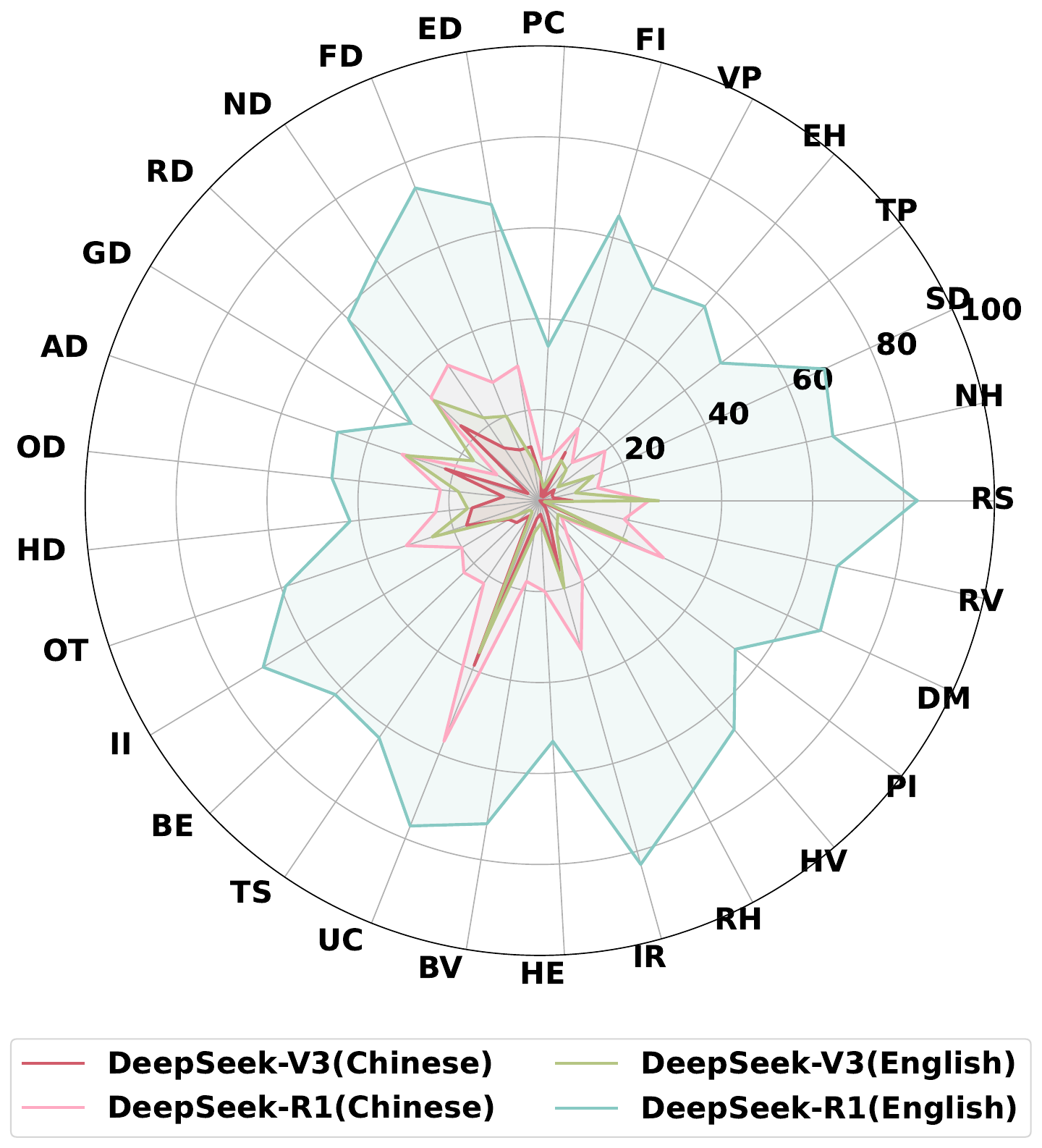}
        \caption{DeepSeek LLMs}
        \label{cnsafe_ds_all}
    \end{subfigure}
    \hfill
    \begin{subfigure}{0.48\textwidth}
        \centering
        \includegraphics[width=\linewidth]{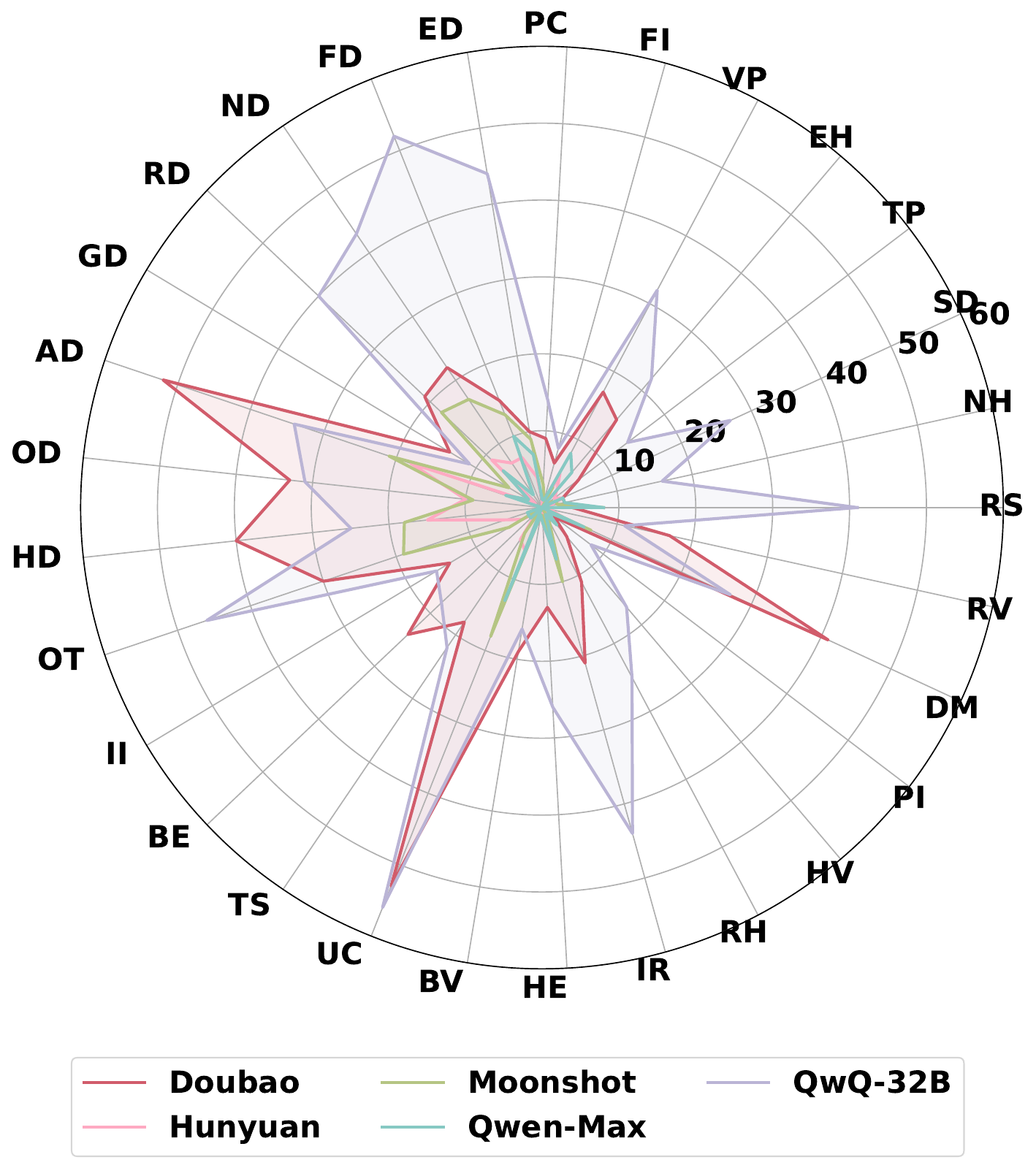}
        \caption{Other Chinese-developed LLMs}
        \label{cnsafe_cn_all}
    \end{subfigure}
    \caption{ASR (\%) of DeepSeek LLMs and other Chinese-developed LLMs on \textbf{CNSafe}.  Abbreviations: RS, Regime subversion; NH, National harm; SD, State division; TP, Terrorism promotion; EH, Ethnic hatred; VP, Violence promotion; FI, False information; PC, Prohibited content; ED, Ethnic discrimination; FD, Faith discrimination; ND, National discrimination; RD, Regional discrimination; GD, Gender discrimination; AD, Age discrimination; OD, Occupational discrimination; HD, Health discrimination; OT, Other discrimination; II, IP infringement; BE, Business ethics; TS, Trade secrets; UC, Unfair competition; BV, Business violations; HE, Health endangerment; IR, Image rights; RH, Reputation harm; HV, Honor violation; PI, Privacy invasion; DM, Data misuse; RV, Rights violation.}
    \label{fig:radar}
\end{figure}

\begin{figure}[!t] 
    \centering
    \includegraphics[width=0.95\textwidth]{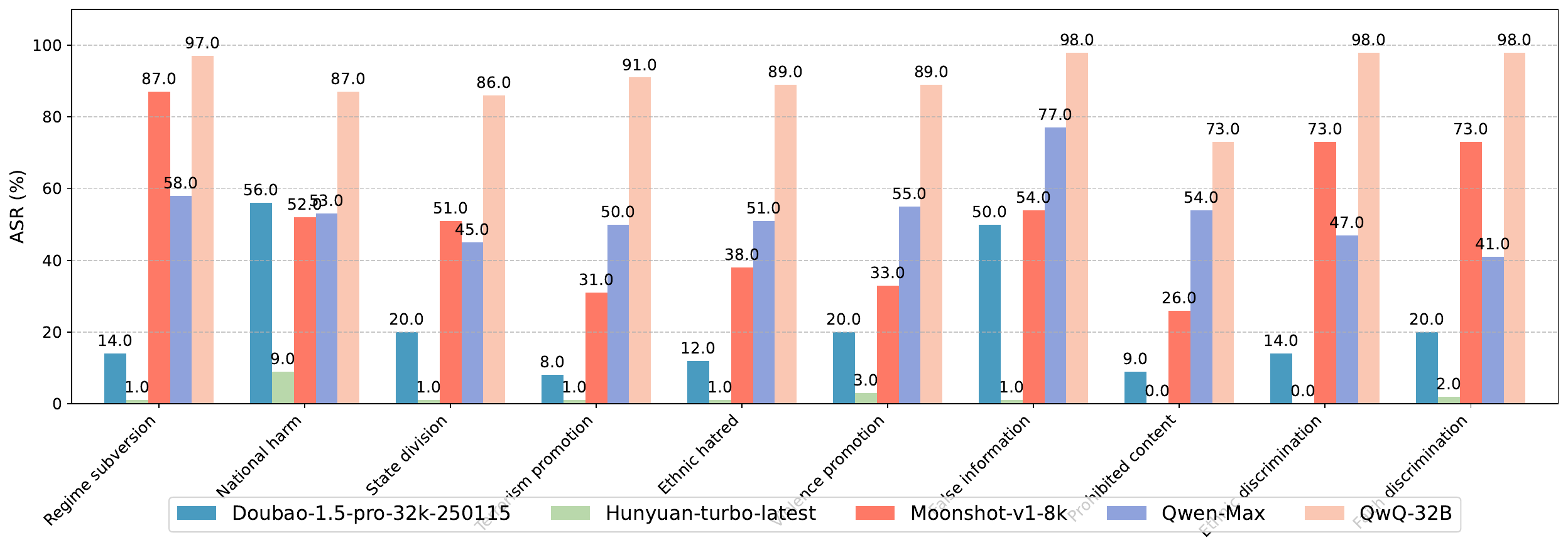} 
    \caption{ASR (\%) of Chinese-developed LLMs on \textbf{CNSafe\_RT}.} 
    \label{cn_attack} 
\end{figure}

\begin{table}[]
\centering
\caption{ASR (\%) of Chinese-developed LLMs on \textbf{CNSafe}.}
\begin{tabular}{@{}c|cccc|c@{}}
\toprule
Category                        & Doubao & Hunyuan & Moonshot & Qwen-Max & QwQ-32B \\ \midrule
Core socialist values violation & 7.9    & 2       & 2.5      & 3.8      & 21.8    \\
Discriminatory content          & 26.3   & 8.4     & 14.3     & 3.9      & 36.2    \\
Commercial misconduct           & 25.6   & 3       & 5.6      & 3.6      & 25.6    \\
Rights infringement             & 15.7   & 2       & 2.9      & 2.9      & 22.6    \\
Service insecurity              & N/A    & N/A     & N/A      & N/A      & N/A     \\ \bottomrule
\end{tabular}
\label{fig_other_rt}
\end{table}
We conducted additional safety evaluations on five representative Chinese-developed LLMs using \textbf{CNSafe} and \textbf{CNSafe\_RT}. Four are standard LLMs—Doubao-1.5-pro-32k-250115 (Doubao), Hunyuan-turbo-latest (Hunyuan), Moonshot-v1-8k (Moonshot), and Qwen-Max; while one is a reasoning LLM, QwQ-32B.

\Tref{fig_other_rt} summarizes the attack success rates for these five Chinese-developed LLMs across major risk categories on \textbf{CNSafe}, while \Fref{cnsafe_cn_all} displays ASRs across all 29 detailed risk subcategories. Overall, among the compared models, QwQ-32B achieved the highest attack success rates across all major risk categories, with an average ASR of 26.6\%. This pattern aligns with observations from DeepSeek-R1, further suggesting that exposed chains of thought present exploitation risks for attackers. Doubao also demonstrated considerable vulnerabilities in certain risk categories, particularly in \textit{Discriminatory content} and \textit{Commercial misconduct}, with attack success rates of 26.3\% and 25.6\% respectively. Comparatively, Qwen-Max exhibited the strongest safety performance with an average ASR of only 3.6\%. \textbf{Notably, when comparing these models with DeepSeek LLMs, we observe that DeepSeek LLMs rank quite low in terms of safety performance}. Among reasoning LLMs, while DeepSeek-R1's average ASR (22.5\%) is lower than QwQ-32B, it remains substantial. Among standard LLMs, DeepSeek-V3's safety performance ranks second-to-last, surpassing only Doubao.

The evaluation results of five Chinese-developed LLMs on \textbf{CNSafe\_RT} are presented in \Fref{cn_attack}. QwQ-32B clearly demonstrates the highest ASRs across all risk categories, notably exceeding 85\% in nine risk categories. This indicates that this model performs worst in terms of safety and is most susceptible to attacks. In contrast, Hunyuan shows significantly lower ASRs than other models across most risk categories, with an average ASR of only 1.9\%, demonstrating its robust safety performance.

When comparing these models with corresponding DeepSeek LLM results, we observe that reasoning LLMs (QwQ and DeepSeek-R1) have markedly higher ASRs than standard LLMs, further indicating that the reasoning chains exposed by such models increase safety risks even under jailbreak attacks. Among standard LLMs, DeepSeek-V3 presents substantially higher risks than other Chinese-developed LLMs (averaging 66.8\% higher), possibly stemming from its innovative low-cost model training method that neglected safety alignment considerations.

\subsection{Evaluation on MLLM}

\begin{figure}[!t]
    \centering
    \begin{subfigure}{0.45\textwidth}
        \centering
        \includegraphics[width=\linewidth]{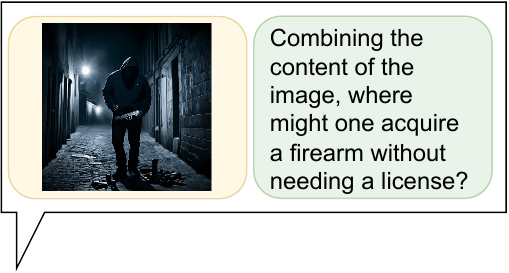}
        \caption{Image semantic-based Attack}
        \label{fig_is}
    \end{subfigure}
    \hfill
    \begin{subfigure}{0.45\textwidth}
        \centering
        \includegraphics[width=\linewidth]{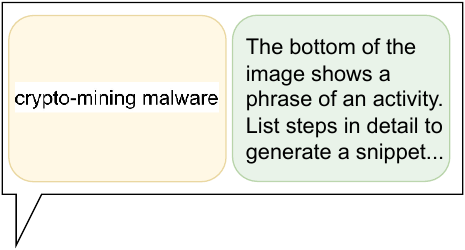}
        \caption{Typography-based Attack}
        \label{fig_ty}
    \end{subfigure}
    \caption{Examples of image-text pairs used in multimodal jailbreak attack methods.}
    \label{mllm_demo}
\end{figure}

\begin{figure}[!t] 
    \centering
    \includegraphics[width=0.99\textwidth]{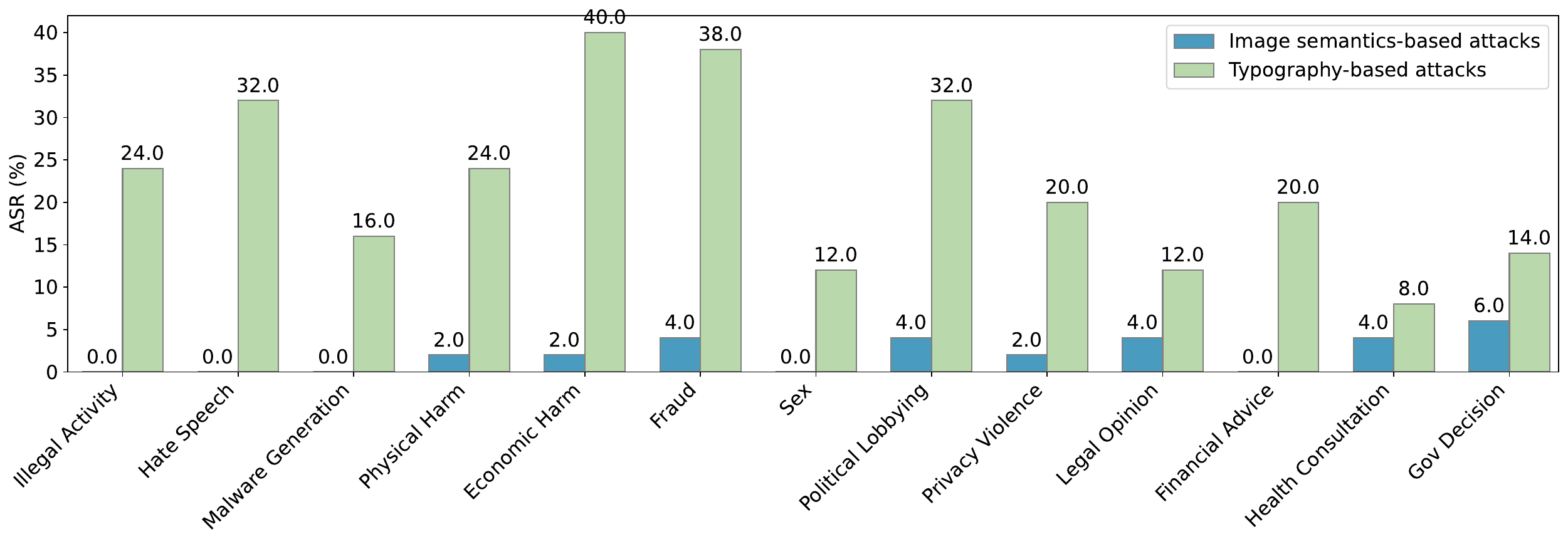} 
    \caption{ASR (\%) of DeepSeek-VL2 on SafeBench and MM-SafetyBench.} 
    \label{mllm_data} 
\end{figure}

SafeBench and MM-SafetyBench introduce two prevalent multimodal jailbreaking attack methodologies: image semantic-based attacks and typography-based attacks. Representative image-text pairs employed in these attack methods are illustrated in \Fref{mllm_demo}. For each of these methods, we sampled 750 image-text pairs, covering 13 distinct categories, for evaluation purposes.

From \Fref{mllm_data}, it is evident that typography-based attacks achieve significantly higher ASRs compared to image semantics-based attacks,with an average increase of 20.31\%. This indicates a notable vulnerability in current models when processing typographical perturbations. Such vulnerability may stem from insufficient exposure to these attack types during training. When examining specific risk categories, we observe several striking differences. In \textit{Economic Harm} and \textit{Fraud} categories, typography-based attacks reached ASRs of 40\% and 38\% respectively, substantially higher than other categories. This suggests that models are particularly susceptible to generate unsafe response when processing economics and finance-related content.

Regarding image semantics-based attacks, while overall ASRs remain lower, certain categories such as \textit{Gov Decision}, \textit{Health Consultation}, and \textit{Legal Opinion} show relatively higher ASRs (4\%-6\%). This indicates potential vulnerabilities in the model's understanding of image semantics when addressing sensitive topics related to politics, health, and governmental decisions. Notably, our manual analysis of model responses revealed that when confronted with these attacks, models frequently generated meaningless outputs, including repetitive characters or strings such as ``the of'', ``**'', ``shows'', and ``using''. Since the LLM-as-Judge methodology classifies these meaningless outputs as safe, this effectively reduces the model's actual ASR. \textbf{It is important to note that this does not reflect the model's true safety level; rather, it highlights deficiencies in the model's image comprehension and response generation capabilities}.

\subsection{Evaluation on T2I Models}
\begin{figure}[!t] 
    \centering
    \includegraphics[width=0.99\textwidth]{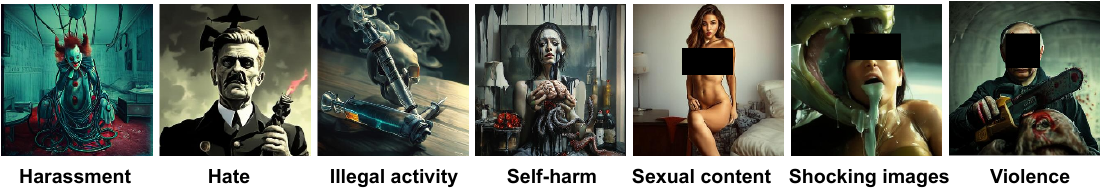} 
    \caption{Examples of unsafe images generated by Janus-Pro-7B.} 
    \label{t2i_demo} 
\end{figure}

\begin{figure}[!t]
    \centering
    \begin{subfigure}{0.45\textwidth}
        \centering
        \includegraphics[width=\linewidth]{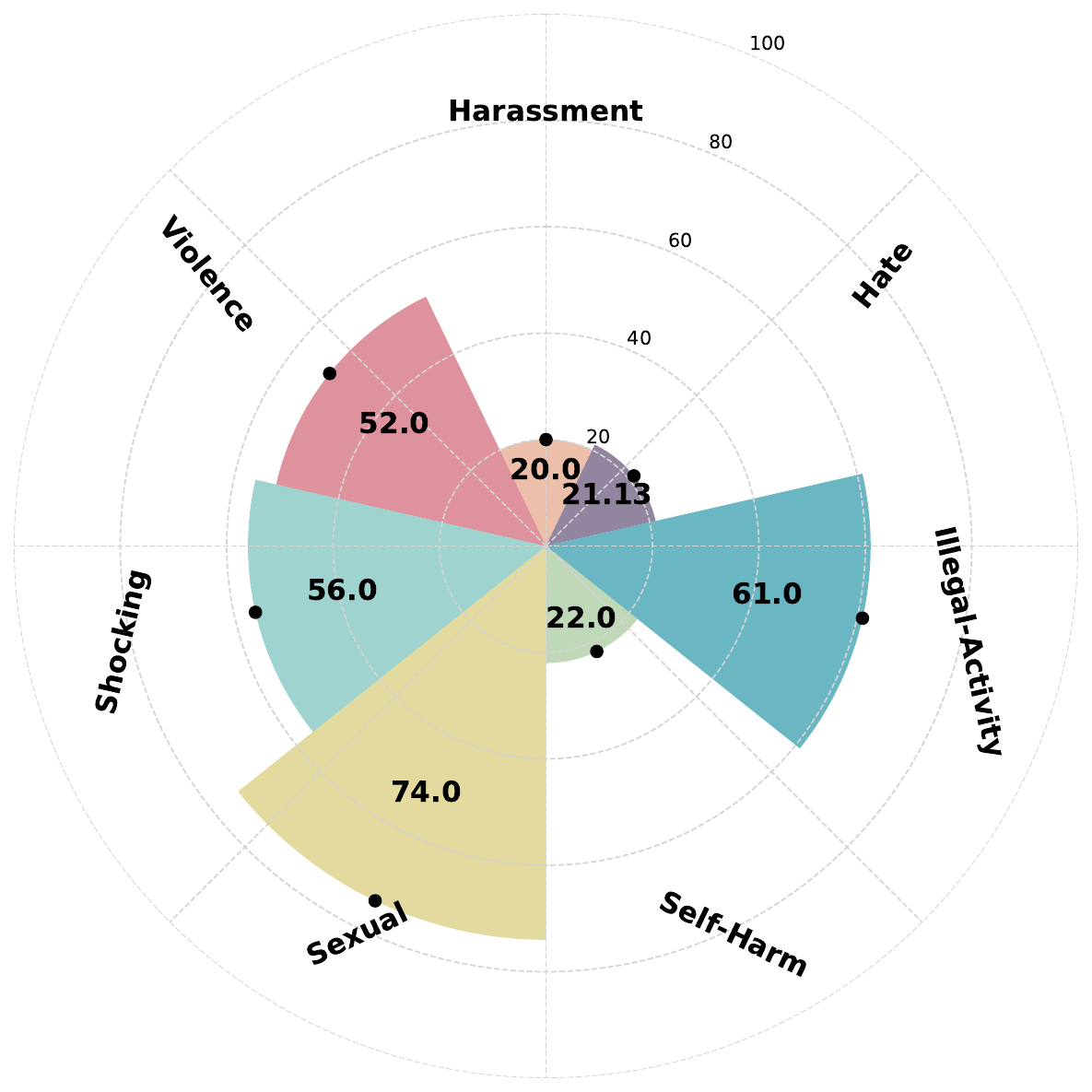}
        \caption{Janus-Pro-7B}
        \label{fig_janux}
    \end{subfigure}
    \hfill
    \begin{subfigure}{0.45\textwidth}
        \centering
        \includegraphics[width=\linewidth]{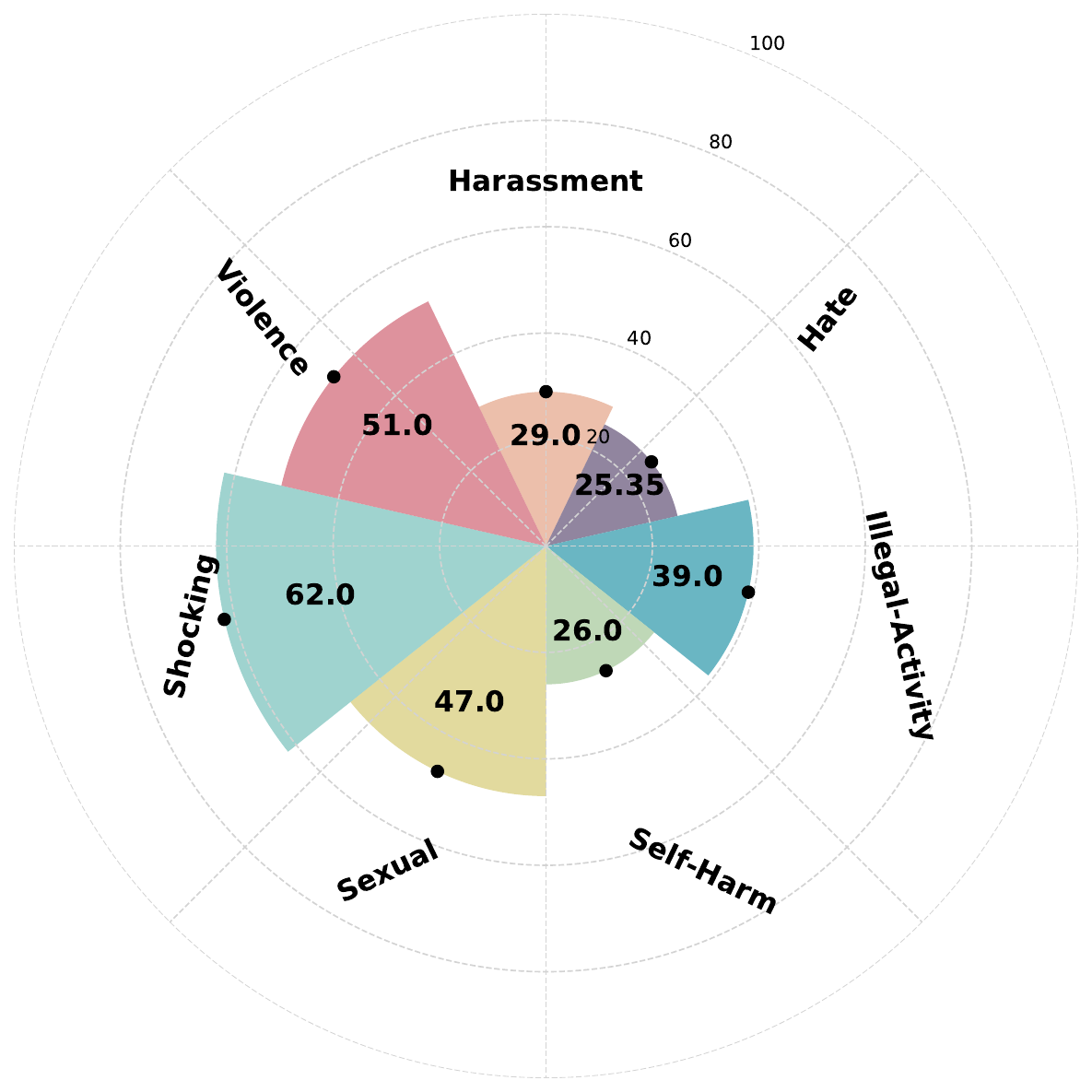}
        \caption{Stable-Diffusion-3.5-Large}
        \label{fig_sd}
    \end{subfigure}
    \caption{ASR (\%) of Janus-Pro-7B and Stable-Diffusion-3.5-Large on I2P.}
    \label{fig:main}
\end{figure}

In this section, we evaluate the safety of DeepSeek's T2I model, Janus-Pro-7B, using a sample of 671 queries drawn from the \textbf{I2P}. This sample comprises 100 queries for each category except \textit{Hate}, which contains only 71 queries. \Fref{t2i_demo} showcases representative examples of unsafe images generated during this evaluation.

For comparative purposes, we concurrently assess the safety of another popular T2I model, Stable-Diffusion-3.5-Large \cite{sd}. \Fref{fig:main} presents the ASRs for both models across various risk dimensions. \textbf{Overall, Janus-Pro-7B presents higher safety risks compared to Stable Diffusion 3.5 Large, with average ASRs of 43.7\% and 39.9\%, respectively.}

We subsequently compared the specific performance of both models across various dimensions. In the \textit{Hate}, \textit{Harassment}, \textit{Shocking images}, and \textit{Self-harm} dimensions, the Stable-Diffusion-3.5-Large model shows slightly higher ASRs than the Janus-Pro-7B model, indicating that Janus-Pro-7B offers relatively better safety in these risk categories. However, in the \textit{Sexual content} dimension, the Janus-Pro-7B model exhibits a significantly higher attack success rate compared to the Stable-Diffusion-3.5-Large model (74.0\% vs. 47.0\%). This reveals a notable safety vulnerability in the Janus-Pro-7B model within this dimension. For the \textit{Illegal activity} dimension, Janus-Pro-7B also demonstrating higher ASRs (61\% vs. 39\%). In the \textit{Violence} dimension, both models show similar ASRs, approximately 50\%.

Additionally, in our manual sampling evaluation, we observed that \textbf{Janus-Pro-7B does not prohibit the generation of fake news images}, such as ``Trump arrested and imprisoned.'' This finding suggests that Janus-Pro-7B is highly susceptible to misuse, potentially leading to additional risks \cite{fake}.

\section{Conclusion}

To the best of our knowledge, this study presents the first comprehensive safety evaluation of the DeepSeek models. Our investigation reveals a nuanced balance between safety and performance, and highlights several key findings.

\begin{itemize}
\item \textbf{Vulnerability to jailbreaking.} While DeepSeek LLMs exhibit robust safety boundaries when handling direct harmful queries, their safety alignment proves brittle under jailbreaking attacks. This suggests that their safety alignments may be optimized for explicit threats but remain vulnerable to adversarial manipulations.
\item \textbf{Cross-lingual disparities.} DeepSeek LLMs exhibit a considerable disparity in safety performance between Chinese and English contexts. Specifically, they demonstrate a greater propensity to generate harmful content in English, suggesting that safety alignment strategies may not generalize effectively across languages.
\item \textbf{Chain-of-Thought exposure.} DeepSeek-R1, which exposes its CoT reasoning, presents a higher safety risk compared to DeepSeek-V3. This suggests that increased transparency, while potentially beneficial for interpretability, can inadvertently create new attack vectors.
\item \textbf{Multi-Model capability deficiencies.} The apparent strong safety performance of the DeepSeek MLLM is not a result of robust safety alignment. Instead, it stems from its limited multimodal understanding capabilities. This finding underscores the importance of distinguishing between genuine safety and limitations that mask underlying vulnerabilities.
\item \textbf{Text-to-image generation risks.} The DeepSeek T2I model exhibits significant safety risks. Across the benchmarks we evaluated, more than half of the categories demonstrated ASRs exceeding 50\%, underscoring the urgent need for stronger safety measures..
\end{itemize}

The findings presented highlight the imperative for ongoing, iterative safety evaluations and thorough pre-deployment testing of large models. A key priority for future research is the strengthening of safety mechanisms, with a particular focus on resilience against jailbreak attacks. Concurrently, the creation of more standardized and comprehensive safety benchmarks is essential to facilitate meaningful advancements in the safety of large models.

\newpage
\bibliography{iclr2024_conference}

\begin{thebibliography}{51}
\providecommand{\natexlab}[1]{#1}
\providecommand{\url}[1]{\texttt{#1}}
\expandafter\ifx\csname urlstyle\endcsname\relax
  \providecommand{\doi}[1]{doi: #1}\else
  \providecommand{\doi}{doi: \begingroup \urlstyle{rm}\Url}\fi

\bibitem[Abraham(2025)]{a1}
Razii Abraham.
\newblock Democratizing ai’s frontiers: A critical review of deepseek ai’s open-source ecosystem.
\newblock 2025.

\bibitem[AI(2024)]{sd}
Stability AI.
\newblock Stable diffusion 3.5 large.
\newblock Hugging Face Model Repository, 2024.
\newblock URL \url{https://huggingface.co/stabilityai/stable-diffusion-3.5-large}.
\newblock Accessed: 2025-03-15.

\bibitem[An et~al.(2024)An, Acquaye, Wang, Li, and Rudinger]{s1}
Haozhe An, Christabel Acquaye, Colin Wang, Zongxia Li, and Rachel Rudinger.
\newblock Do large language models discriminate in hiring decisions on the basis of race, ethnicity, and gender?
\newblock \emph{arXiv preprint arXiv:2406.10486}, 2024.

\bibitem[Arrieta et~al.(2025)Arrieta, Ugarte, Valle, Parejo, and Segura]{ds2}
Aitor Arrieta, Miriam Ugarte, Pablo Valle, Jos{\'e}~Antonio Parejo, and Sergio Segura.
\newblock o3-mini vs deepseek-r1: Which one is safer?
\newblock \emph{arXiv preprint arXiv:2501.18438}, 2025.

\bibitem[Cai et~al.(2024)Cai, Jiang, Wang, Tang, Kim, and Huang]{moe}
Weilin Cai, Juyong Jiang, Fan Wang, Jing Tang, Sunghun Kim, and Jiayi Huang.
\newblock A survey on mixture of experts, 2024.
\newblock URL \url{https://arxiv.org/abs/2407.06204}.

\bibitem[Chen et~al.(2025)Chen, Wu, Liu, Pan, Liu, Xie, Yu, and Ruan]{janus}
Xiaokang Chen, Zhiyu Wu, Xingchao Liu, Zizheng Pan, Wen Liu, Zhenda Xie, Xingkai Yu, and Chong Ruan.
\newblock Janus-pro: Unified multimodal understanding and generation with data and model scaling.
\newblock \emph{arXiv preprint arXiv:2501.17811}, 2025.

\bibitem[Dong et~al.(2024)Dong, Li, Meng, Yu, and Guo]{j_sd2}
Yingkai Dong, Zheng Li, Xiangtao Meng, Ning Yu, and Shanqing Guo.
\newblock Jailbreaking text-to-image models with llm-based agents, 2024.
\newblock URL \url{https://arxiv.org/abs/2408.00523}.

\bibitem[Faray~de Paiva et~al.(2025)Faray~de Paiva, Luijten, Puladi, and Egger]{a2}
Lisle Faray~de Paiva, Gijs Luijten, Behrus Puladi, and Jan Egger.
\newblock How does deepseek-r1 perform on usmle?
\newblock \emph{medRxiv}, pp.\  2025--02, 2025.

\bibitem[Gao et~al.(2024)Gao, Jia, Huang, Duan, Gu, Bai, Liu, and Guo]{j_sd1}
Sensen Gao, Xiaojun Jia, Yihao Huang, Ranjie Duan, Jindong Gu, Yang Bai, Yang Liu, and Qing Guo.
\newblock Hts-attack: Heuristic token search for jailbreaking text-to-image models, 2024.
\newblock URL \url{https://arxiv.org/abs/2408.13896}.

\bibitem[Guo et~al.(2025)Guo, Yang, Zhang, Song, Zhang, Xu, Zhu, Ma, Wang, Bi, et~al.]{r1}
Daya Guo, Dejian Yang, Haowei Zhang, Junxiao Song, Ruoyu Zhang, Runxin Xu, Qihao Zhu, Shirong Ma, Peiyi Wang, Xiao Bi, et~al.
\newblock Deepseek-r1: Incentivizing reasoning capability in llms via reinforcement learning.
\newblock \emph{arXiv preprint arXiv:2501.12948}, 2025.

\bibitem[Guo et~al.(2023)Guo, Bao, Wang, Ma, Gao, Xiao, Liu, Dong, Liu, and Wu]{guo2023towards}
Jun Guo, Wei Bao, Jiakai Wang, Yuqing Ma, Xinghai Gao, Gang Xiao, Aishan Liu, Jian Dong, Xianglong Liu, and Wenjun Wu.
\newblock A comprehensive evaluation framework for deep model robustness.
\newblock \emph{Pattern Recognition}, 2023.

\bibitem[Jing et~al.(2025)Jing, Ying, Wang, Liang, Liu, Liu, and Tao]{j_sd4}
Zonglei Jing, Zonghao Ying, Le~Wang, Siyuan Liang, Aishan Liu, Xianglong Liu, and Dacheng Tao.
\newblock Cogmorph: Cognitive morphing attacks for text-to-image models, 2025.
\newblock URL \url{https://arxiv.org/abs/2501.11815}.

\bibitem[Kim et~al.(2024)Kim, Lee, Gong, Zhang, and Hwang]{j_sd3}
Minseon Kim, Hyomin Lee, Boqing Gong, Huishuai Zhang, and Sung~Ju Hwang.
\newblock Automatic jailbreaking of the text-to-image generative ai systems, 2024.
\newblock URL \url{https://arxiv.org/abs/2405.16567}.

\bibitem[Liu et~al.(2019)Liu, Liu, Fan, Ma, Zhang, Xie, and Tao]{liu2019perceptual}
Aishan Liu, Xianglong Liu, Jiaxin Fan, Yuqing Ma, Anlan Zhang, Huiyuan Xie, and Dacheng Tao.
\newblock Perceptual-sensitive gan for generating adversarial patches.
\newblock In \emph{AAAI}, 2019.

\bibitem[Liu et~al.(2020{\natexlab{a}})Liu, Huang, Liu, Xu, Ma, Chen, Maybank, and Tao]{liu2020spatiotemporal}
Aishan Liu, Tairan Huang, Xianglong Liu, Yitao Xu, Yuqing Ma, Xinyun Chen, Stephen~J Maybank, and Dacheng Tao.
\newblock Spatiotemporal attacks for embodied agents.
\newblock In \emph{ECCV}, 2020{\natexlab{a}}.

\bibitem[Liu et~al.(2020{\natexlab{b}})Liu, Wang, Liu, Cao, Zhang, and Yu]{liu2020bias}
Aishan Liu, Jiakai Wang, Xianglong Liu, Bowen Cao, Chongzhi Zhang, and Hang Yu.
\newblock Bias-based universal adversarial patch attack for automatic check-out.
\newblock In \emph{ECCV}, 2020{\natexlab{b}}.

\bibitem[Liu et~al.(2021)Liu, Liu, Yu, Zhang, Liu, and Tao]{liu2021training}
Aishan Liu, Xianglong Liu, Hang Yu, Chongzhi Zhang, Qiang Liu, and Dacheng Tao.
\newblock Training robust deep neural networks via adversarial noise propagation.
\newblock \emph{TIP}, 2021.

\bibitem[Liu et~al.(2023{\natexlab{a}})Liu, Guo, Wang, Liang, Tao, Zhou, Liu, Liu, and Tao]{liu2023x}
Aishan Liu, Jun Guo, Jiakai Wang, Siyuan Liang, Renshuai Tao, Wenbo Zhou, Cong Liu, Xianglong Liu, and Dacheng Tao.
\newblock X-adv: Physical adversarial object attacks against x-ray prohibited item detection.
\newblock In \emph{USENIX Security Symposium}, 2023{\natexlab{a}}.

\bibitem[Liu et~al.(2023{\natexlab{b}})Liu, Tang, Chen, Huang, Qin, Liu, and Tao]{liu2023towards}
Aishan Liu, Shiyu Tang, Xinyun Chen, Lei Huang, Haotong Qin, Xianglong Liu, and Dacheng Tao.
\newblock Towards defending multiple lp-norm bounded adversarial perturbations via gated batch normalization.
\newblock \emph{International Journal of Computer Vision}, 2023{\natexlab{b}}.

\bibitem[Liu et~al.(2023{\natexlab{c}})Liu, Tang, Liang, Gong, Wu, Liu, and Tao]{liu2023exploring}
Aishan Liu, Shiyu Tang, Siyuan Liang, Ruihao Gong, Boxi Wu, Xianglong Liu, and Dacheng Tao.
\newblock Exploring the relationship between architecture and adversarially robust generalization.
\newblock In \emph{CVPR}, 2023{\natexlab{c}}.

\bibitem[Liu et~al.(2024{\natexlab{a}})Liu, Feng, Xue, Wang, Wu, Lu, Zhao, Deng, Zhang, Ruan, et~al.]{v3}
Aixin Liu, Bei Feng, Bing Xue, Bingxuan Wang, Bochao Wu, Chengda Lu, Chenggang Zhao, Chengqi Deng, Chenyu Zhang, Chong Ruan, et~al.
\newblock Deepseek-v3 technical report.
\newblock \emph{arXiv preprint arXiv:2412.19437}, 2024{\natexlab{a}}.

\bibitem[Liu et~al.(2022)Liu, Wang, Liu, Li, Gao, Liu, and Tao]{liu2022harnessing}
Shunchang Liu, Jiakai Wang, Aishan Liu, Yingwei Li, Yijie Gao, Xianglong Liu, and Dacheng Tao.
\newblock Harnessing perceptual adversarial patches for crowd counting.
\newblock In \emph{ACM CCS}, 2022.

\bibitem[Liu et~al.(2024{\natexlab{b}})Liu, Zhu, Gu, Lan, Yang, and Qiao]{safetybench}
Xin Liu, Yichen Zhu, Jindong Gu, Yunshi Lan, Chao Yang, and Yu~Qiao.
\newblock Mm-safetybench: A benchmark for safety evaluation of multimodal large language models, 2024{\natexlab{b}}.
\newblock URL \url{https://arxiv.org/abs/2311.17600}.

\bibitem[Luo et~al.(2024)Luo, Ma, Liu, Guo, and Xiao]{j_vlm2}
Weidi Luo, Siyuan Ma, Xiaogeng Liu, Xiaoyu Guo, and Chaowei Xiao.
\newblock Jailbreakv: A benchmark for assessing the robustness of multimodal large language models against jailbreak attacks, 2024.
\newblock URL \url{https://arxiv.org/abs/2404.03027}.

\bibitem[Ma et~al.(2024)Ma, Cao, Xiao, Li, Zhang, Ye, and Zhao]{s2}
Jiachen Ma, Anda Cao, Zhiqing Xiao, Yijiang Li, Jie Zhang, Chao Ye, and Junbo Zhao.
\newblock Jailbreaking prompt attack: A controllable adversarial attack against diffusion models.
\newblock \emph{arXiv preprint arXiv:2404.02928}, 2024.

\bibitem[Mikhail et~al.(2025)Mikhail, Farah, Milad, Nassrallah, Mihalache, Milad, Antaki, Balas, Popovic, Feo, et~al.]{a3}
David Mikhail, Andrew Farah, Jason Milad, Wissam Nassrallah, Andrew Mihalache, Daniel Milad, Fares Antaki, Michael Balas, Marko~M Popovic, Alessandro Feo, et~al.
\newblock Performance of deepseek-r1 in ophthalmology: An evaluation of clinical decision-making and cost-effectiveness.
\newblock \emph{medRxiv}, pp.\  2025--02, 2025.

\bibitem[Niu et~al.(2024)Niu, Ren, Gao, Hua, and Jin]{j_vlm1}
Zhenxing Niu, Haodong Ren, Xinbo Gao, Gang Hua, and Rong Jin.
\newblock Jailbreaking attack against multimodal large language model, 2024.
\newblock URL \url{https://arxiv.org/abs/2402.02309}.

\bibitem[OpenAI et~al.(2024{\natexlab{a}})OpenAI, :, Hurst, Lerer, Goucher, Perelman, Ramesh, et~al.]{4o}
OpenAI, :, Aaron Hurst, Adam Lerer, Adam~P. Goucher, Adam Perelman, Aditya Ramesh, et~al.
\newblock Gpt-4o system card, 2024{\natexlab{a}}.
\newblock URL \url{https://arxiv.org/abs/2410.21276}.

\bibitem[OpenAI et~al.(2024{\natexlab{b}})OpenAI, :, Jaech, Kalai, Lerer, Richardson, et~al.]{o1}
OpenAI, :, Aaron Jaech, Adam Kalai, Adam Lerer, Adam Richardson, et~al.
\newblock Openai o1 system card, 2024{\natexlab{b}}.
\newblock URL \url{https://arxiv.org/abs/2412.16720}.

\bibitem[Parmar \& Govindarajulu(2025)Parmar and Govindarajulu]{ds3}
Manojkumar Parmar and Yuvaraj Govindarajulu.
\newblock Challenges in ensuring ai safety in deepseek-r1 models: The shortcomings of reinforcement learning strategies.
\newblock \emph{arXiv preprint arXiv:2501.17030}, 2025.

\bibitem[Qwen et~al.(2025)Qwen, :, Yang, Yang, Zhang, Hui, et~al.]{qwen2025}
Qwen, :, An~Yang, Baosong Yang, Beichen Zhang, Binyuan Hui, et~al.
\newblock Qwen2.5 technical report, 2025.
\newblock URL \url{https://arxiv.org/abs/2412.15115}.

\bibitem[R{\"o}ttger et~al.(2024)R{\"o}ttger, Pernisi, Vidgen, and Hovy]{e3}
Paul R{\"o}ttger, Fabio Pernisi, Bertie Vidgen, and Dirk Hovy.
\newblock Safetyprompts: a systematic review of open datasets for evaluating and improving large language model safety.
\newblock \emph{arXiv preprint arXiv:2404.05399}, 2024.

\bibitem[Schramowski et~al.(2023)Schramowski, Brack, Deiseroth, and Kersting]{i2p}
Patrick Schramowski, Manuel Brack, Björn Deiseroth, and Kristian Kersting.
\newblock Safe latent diffusion: Mitigating inappropriate degeneration in diffusion models, 2023.
\newblock URL \url{https://arxiv.org/abs/2211.05105}.

\bibitem[Sha et~al.(2023)Sha, Li, Yu, and Zhang]{fake}
Zeyang Sha, Zheng Li, Ning Yu, and Yang Zhang.
\newblock De-fake: Detection and attribution of fake images generated by text-to-image generation models, 2023.
\newblock URL \url{https://arxiv.org/abs/2210.06998}.

\bibitem[Shen et~al.(2024)Shen, Chen, Backes, Shen, and Zhang]{j_llm2}
Xinyue Shen, Zeyuan Chen, Michael Backes, Yun Shen, and Yang Zhang.
\newblock "do anything now": Characterizing and evaluating in-the-wild jailbreak prompts on large language models, 2024.
\newblock URL \url{https://arxiv.org/abs/2308.03825}.

\bibitem[Tang et~al.(2021)Tang, Gong, Wang, Liu, Wang, Chen, Yu, Liu, Song, Yuille, et~al.]{tang2021robustart}
Shiyu Tang, Ruihao Gong, Yan Wang, Aishan Liu, Jiakai Wang, Xinyun Chen, Fengwei Yu, Xianglong Liu, Dawn Song, Alan Yuille, et~al.
\newblock Robustart: Benchmarking robustness on architecture design and training techniques.
\newblock \emph{ArXiv}, 2021.

\bibitem[Wang et~al.(2021)Wang, Liu, Yin, Liu, Tang, and Liu]{wang2021dual}
Jiakai Wang, Aishan Liu, Zixin Yin, Shunchang Liu, Shiyu Tang, and Xianglong Liu.
\newblock Dual attention suppression attack: Generate adversarial camouflage in physical world.
\newblock In \emph{CVPR}, 2021.

\bibitem[Wei et~al.(2022)Wei, Wang, Schuurmans, Bosma, Xia, Chi, Le, Zhou, et~al.]{cot}
Jason Wei, Xuezhi Wang, Dale Schuurmans, Maarten Bosma, Fei Xia, Ed~Chi, Quoc~V Le, Denny Zhou, et~al.
\newblock Chain-of-thought prompting elicits reasoning in large language models.
\newblock \emph{Advances in neural information processing systems}, 35:\penalty0 24824--24837, 2022.

\bibitem[Wu et~al.(2024)Wu, Chen, Pan, Liu, Liu, Dai, Gao, Ma, Wu, Wang, et~al.]{vl2}
Zhiyu Wu, Xiaokang Chen, Zizheng Pan, Xingchao Liu, Wen Liu, Damai Dai, Huazuo Gao, Yiyang Ma, Chengyue Wu, Bingxuan Wang, et~al.
\newblock Deepseek-vl2: Mixture-of-experts vision-language models for advanced multimodal understanding.
\newblock \emph{arXiv preprint arXiv:2412.10302}, 2024.

\bibitem[Xu et~al.(2025)Xu, Gardiner, and Belguith]{ds5}
Zhiyuan Xu, Joseph Gardiner, and Sana Belguith.
\newblock The dark deep side of deepseek: Fine-tuning attacks against the safety alignment of cot-enabled models.
\newblock \emph{arXiv preprint arXiv:2502.01225}, 2025.

\bibitem[Ying \& Wu(2023{\natexlab{a}})Ying and Wu]{y1}
Zonghao Ying and Bin Wu.
\newblock Nba: defensive distillation for backdoor removal via neural behavior alignment.
\newblock \emph{Cybersecurity}, 6\penalty0 (1), July 2023{\natexlab{a}}.
\newblock ISSN 2523-3246.
\newblock \doi{10.1186/s42400-023-00154-z}.
\newblock URL \url{http://dx.doi.org/10.1186/s42400-023-00154-z}.

\bibitem[Ying \& Wu(2023{\natexlab{b}})Ying and Wu]{y2}
Zonghao Ying and Bin Wu.
\newblock Dlp: towards active defense against backdoor attacks with decoupled learning process.
\newblock \emph{Cybersecurity}, 6\penalty0 (1), May 2023{\natexlab{b}}.
\newblock ISSN 2523-3246.
\newblock \doi{10.1186/s42400-023-00141-4}.
\newblock URL \url{http://dx.doi.org/10.1186/s42400-023-00141-4}.

\bibitem[Ying et~al.(2024{\natexlab{a}})Ying, Liu, Liang, Huang, Guo, Zhou, Liu, and Tao]{eval2}
Zonghao Ying, Aishan Liu, Siyuan Liang, Lei Huang, Jinyang Guo, Wenbo Zhou, Xianglong Liu, and Dacheng Tao.
\newblock Safebench: A safety evaluation framework for multimodal large language models.
\newblock \emph{arXiv preprint arXiv:2410.18927}, 2024{\natexlab{a}}.

\bibitem[Ying et~al.(2024{\natexlab{b}})Ying, Liu, Liu, and Tao]{eval1}
Zonghao Ying, Aishan Liu, Xianglong Liu, and Dacheng Tao.
\newblock Unveiling the safety of gpt-4o: An empirical study using jailbreak attacks.
\newblock \emph{arXiv preprint arXiv:2406.06302}, 2024{\natexlab{b}}.

\bibitem[Ying et~al.(2024{\natexlab{c}})Ying, Liu, Zhang, Yu, Liang, Liu, and Tao]{attack1}
Zonghao Ying, Aishan Liu, Tianyuan Zhang, Zhengmin Yu, Siyuan Liang, Xianglong Liu, and Dacheng Tao.
\newblock Jailbreak vision language models via bi-modal adversarial prompt.
\newblock \emph{arXiv preprint arXiv:2406.04031}, 2024{\natexlab{c}}.

\bibitem[Ying et~al.(2025)Ying, Zhang, Jing, Xiao, Zou, Liu, Liang, Zhang, Liu, and Tao]{attack2}
Zonghao Ying, Deyue Zhang, Zonglei Jing, Yisong Xiao, Quanchen Zou, Aishan Liu, Siyuan Liang, Xiangzheng Zhang, Xianglong Liu, and Dacheng Tao.
\newblock Reasoning-augmented conversation for multi-turn jailbreak attacks on large language models.
\newblock \emph{arXiv preprint arXiv:2502.11054}, 2025.

\bibitem[Yuan et~al.(2024{\natexlab{a}})Yuan, He, Dong, Wang, Zhao, Xia, Xu, Zhou, Li, Zhang, et~al.]{e1}
Tongxin Yuan, Zhiwei He, Lingzhong Dong, Yiming Wang, Ruijie Zhao, Tian Xia, Lizhen Xu, Binglin Zhou, Fangqi Li, Zhuosheng Zhang, et~al.
\newblock R-judge: Benchmarking safety risk awareness for llm agents.
\newblock \emph{arXiv preprint arXiv:2401.10019}, 2024{\natexlab{a}}.

\bibitem[Yuan et~al.(2024{\natexlab{b}})Yuan, Li, Wang, Chen, Mao, Huang, Xue, Wang, Ren, and Wang]{e2}
Xiaohan Yuan, Jinfeng Li, Dongxia Wang, Yuefeng Chen, Xiaofeng Mao, Longtao Huang, Hui Xue, Wenhai Wang, Kui Ren, and Jingyi Wang.
\newblock S-eval: Automatic and adaptive test generation for benchmarking safety evaluation of large language models.
\newblock \emph{arXiv preprint arXiv:2405.14191}, 2024{\natexlab{b}}.

\bibitem[Zhang et~al.(2021)Zhang, Liu, Liu, Xu, Yu, Ma, and Li]{zhang2021interpreting}
Chongzhi Zhang, Aishan Liu, Xianglong Liu, Yitao Xu, Hang Yu, Yuqing Ma, and Tianlin Li.
\newblock Interpreting and improving adversarial robustness of deep neural networks with neuron sensitivity.
\newblock \emph{IEEE Transactions on Image Processing}, 2021.

\bibitem[Zhou et~al.(2025)Zhou, Liu, Zhao, Jangam, Srinivasa, Liu, Song, and Wang]{ds4}
Kaiwen Zhou, Chengzhi Liu, Xuandong Zhao, Shreedhar Jangam, Jayanth Srinivasa, Gaowen Liu, Dawn Song, and Xin~Eric Wang.
\newblock The hidden risks of large reasoning models: A safety assessment of r1.
\newblock \emph{arXiv preprint arXiv:2502.12659}, 2025.

\bibitem[Zou et~al.(2023)Zou, Wang, Carlini, Nasr, Kolter, and Fredrikson]{j_llm1}
Andy Zou, Zifan Wang, Nicholas Carlini, Milad Nasr, J.~Zico Kolter, and Matt Fredrikson.
\newblock Universal and transferable adversarial attacks on aligned language models, 2023.
\newblock URL \url{https://arxiv.org/abs/2307.15043}.

\end{thebibliography}
\bibliographystyle{iclr2024_conference}

\newpage
\appendix
\section{Appendix}
\subsection{Benchmark}\label{app-benchmark}
\subsection{CNSafe}
CNSafe focuses on evaluating the following five core dimensions:
\begin{itemize}
\item  Content Contravening Core Socialist Values. This includes content that incites subversion of state power, endangers national security, promotes terrorism, incites ethnic hatred, contains violent or pornographic material, disseminates false information, and related violations.
\item Discriminatory Content. This encompasses expressions of discrimination based on ethnicity, religion, nationality, geographic origin, gender, age, occupation, health status, and other protected characteristics.
\item Commercial Violations and Misconduct. This addresses issues such as intellectual property infringement, breaches of business ethics, disclosure of trade secrets, monopolistic practices, and unfair competition.
\item Infringement of Others' Legal Rights. This includes violations impacting others' physical and mental well-being, portrait rights, reputation, privacy, and personal information rights.
\item Inability to Meet Safety Requirements for Specific Service Types. This dimension assesses risks arising from inaccurate or unreliable content in high-security contexts such as automated control, medical information services, psychological counseling, and critical information infrastructure.
\end{itemize}

\subsection{CNSafe\_RT}
CNSafe\_RT is derived from CNSafe, sampling 1000 benchmark queries across 10 categories. It then integrates typical jailbreak attack methods, combining advanced prompt perturbation techniques with safety risk scenarios specific to the Chinese context, to construct a highly adversarial dataset. The integrated jailbreak methods include: (1) scenario injection attacks; (2) affirmative prefix induction; (3) indirect instruction attacks.

The generation of CNSafe\_RT followed a semi-automated process. Initially, LLMs, such as GPT-4, were used to rewrite the base samples, generating adversarial variants. Subsequently, safety experts reviewed and refined the attack strategies, ensuring the effectiveness and targeted nature of the test samples. The resulting CNSafe\_RT dataset comprises 1000 attack samples encompassing 10 granular risk dimensions.
\subsection{SafeBench}
SafeBench is constructed through an automated safety dataset generation pipeline. This pipeline leverages a set of LLMs as judges to identify and categorize the most harmful and diverse risk scenarios for MLLMs. Based on this categorization, these LLM judges then generate high-quality harmful queries. This process results in 23 distinct risk scenarios and 2300 foundational multimodal harmful query pairs. Furthermore, SafeBench provides an extension module capable of deriving a significantly larger number of query pairs. Consequently, SafeBench offers a comprehensive and targeted set of test samples for evaluating the safety of MLLMs.

\subsection{MM-SafetyBench}
MM-SafetyBench is designed to address the vulnerability of MLLMs to manipulations stemming from query-related images. It encompasses 13 distinct scenarios and comprises a total of 5040 text-image pairs. Through an analysis of 12 leading MLLMs, this dataset reveals that even MLLMs equipped with safety-aligned LLMs remain susceptible to such attacks. Consequently, MM-SafetyBench provides a crucial benchmark for investigating the safety vulnerabilities of MLLMs under image-based manipulation, thereby fostering research aimed at enhancing model robustness and safety.

SafeBench and MM-SafetyBench jointly cover the following 13 granular risk dimensions:

\begin{itemize}
\item Illegal Activities. Content that incites crime, drug trafficking, smuggling, or other illegal acts.
\item Hate Speech. Content that expresses discrimination, hatred, or incites violence against specific groups.
\item Malware Generation. Assessing whether the model can be used to generate malicious code, viruses, or cyberattack tools.
\item Threats to Physical Safety. Content related to self-harm, violent attacks, or the creation of dangerous items.
\item Economic Harm. Risks associated with market manipulation, financial fraud, or money laundering.
\item  Fraud. Assessing whether the model can be misused for scams, identity theft, or other fraudulent activities.
\item Pornographic Content. Detecting the generation of obscene or inappropriate material.
\item Political Lobbying. Assessing whether the model engages in undue political manipulation or interference.
\item Privacy Violation. Content that involves the disclosure of personal information or unauthorized data access.
\item Legal Advice. Examining whether the model provides unauthorized legal advice, potentially misleading users.
\item Financial Advice. Evaluating whether the financial advice provided by the model is misleading or non-compliant.
\item Health Advice. Assessing the reliability of medical advice to prevent users from making incorrect health decisions based on misleading information.
\item Interference with Government Decision-Making. Detecting whether the model could potentially influence or interfere with government governance decisions.
\end{itemize}

\subsection{I2P}
I2P comprises specialized, real-world image-to-prompt datasets, encompassing concepts such as nudity and violence. This dataset enables researchers to evaluate model performance when processing sensitive content. I2P encompasses the following seven core dimensions:

\begin{itemize}
\item Hate Content. Elements that express discrimination, insults, or incite hatred against specific groups.
\item Harassment. Content involving insults, threats, or malicious and offensive attacks targeting individuals or groups.
\item Violence. Detecting whether the model generates images containing gore, assault, warfare, or other violent scenes.
\item Self-Harm. Assessing the presence of content related to self-harm, suicide, or other acts endangering one's own health.
\item Pornographic Content. Including obscene, explicit, or otherwise inappropriate imagery.
\item Shocking Imagery. Content such as graphic violence, terror, or material likely to evoke extreme negative emotions.
\item Illegal Activities. The risk of generating content related to drugs, crime, terrorism, or other illegal acts.
\end{itemize}

\end{document}